\documentclass[twocolumn,a4paper,superscriptaddress,showpacs,prl]{revtex4}
\usepackage{amsmath}
\usepackage{amsfonts}
\usepackage{graphicx}
\usepackage{longtable}
\begin{document}
\title{Generalized information theoretic measure \\ to discern the {\em quantumness} 
of correlations} 
\affiliation{Department of Physics, Bangalore University, 
Bangalore-560 056, India.}\affiliation{Department of Computer Science and Center for Quantum Studies, George 
Mason 
University, Fairfax, VA 22030, USA.}\affiliation{Inspire Institute Inc., McLean, VA 22101, USA.}
\author{A. R. Usha Devi}
\email{arutth@rediffmail.com}
\affiliation{Department of Physics, Bangalore University, 
Bangalore-560 056, India.}  
\affiliation{Inspire Institute Inc., McLean, VA 22101, USA.}
\author{A. K. Rajagopal}
\affiliation{Department of Computer Science and Center for Quantum Studies, George Mason 
University, Fairfax, VA 22030, USA.} 
\affiliation{Inspire Institute Inc., McLean, VA 22101, USA.}

\date{\today}

\begin{abstract} 
A novel measure, quantumness of correlations, ${\cal Q}_{AB}$, is
introduced here for bipartite states, by incorporating the required measurement scheme 
crucial in defining any such quantity. Quantumness coincides with the  previously 
proposed measures in special cases and it vanishes for separable states - a feature not 
captured by the measures proposed earlier. It is found that an optimal generalized measurement  
on one of the parts leaves the overall state in its closest  separable form, which shares the same 
marginal for the other part, implying that  ${\cal Q}_{AB}$ is non-zero for all 
entangled bipartite states and it serves as an upper bound to the relative entropy of entanglement. 
    
\end{abstract}
\maketitle
In classical probability theory, two random variables $A$ and $B$
are said to be correlated if their  probability distribution, $P(a,b)$ 
cannot be expressed as a mere product of the marginal probabilities  
$P(a)$ and $P(b)$.   There are several equivalent quantitative  measures 
of testing these classical correlations~(CC). In the quantum description, 
probability distributions are replaced by  density operators and 
the statement, {\em a bipartite density matrix is correlated, if it cannot 
be expressed  in a simple  product form of its constituent 
density matrices} provides a natural extension of the idea of correlations. 
Several counter-intuitive features raise their heads due to 
{\em quantum correlations}~(QC) exhibited by subsystems of a 
composite state and  have led to philosophical debates in the conceptual 
understanding of quantum theory, following  
Einstein-Podolsky-Rosen criticism~\cite{EPR}. The rapid progress of quantum information
science~\cite{Zol, Horreview} has stimulated intense activity recently
on the characterization of genuine
QC. It may be emphasized here that 
 the notion of correlation  {\it per se} does not set a borderline between   
classical or quantum descriptions. Following Werner~\cite{werner}, 
a bipartite density matrix $\hat{\rho}_{AB}$ is 
{\em classically correlated} or {\em separable} if it admits 
a convex combination of product states as 
\begin{equation}
\label{wersep}
\hat{\rho}^{{\rm (sep)}}_{AB}=\displaystyle\sum_{i}\,p_i\, \hat\rho^{(i)}_A\otimes\hat\rho_B^{(i)},\ \ 0\leq 
p_i\leq 1,\ 
\displaystyle\sum_i\,{p_i}=1.
\end{equation}  
This provides a generalization of the simplest uncorrelated state, 
$\hat\rho_{AB}~=~\hat\rho_A\otimes\hat\rho_B,$ and forms a basis for the characterization 
and quantification of QC that exist over and above the classical ones. 
However, several research groups~\cite{OZ, Ved, RR, Hamieh, ECG} claim   
that separability (absence of quantum entanglement) does not necessarily imply 
CC.  
In particular, Ollivier and Zurek~[OZ]~\cite{OZ} argued that any measure of QC 
must be based on the fact that {\em it is impossible, through measurement on one part of a composite state,
 to ascribe  independent reality to  the other part} and hence, 
 a measurement scheme,  which verifies this basic feature, forms an essential ingredient 
 of this characterization. A bipartite state is classically correlated, if there exists an 
 optimal measurement on a part of the correlated  system, such that the overall  state  
remains unperturbed~\cite{OZ}. It has been found that~\cite{OZ, Ved, RR} 
only a subset of separable states 
show such insensitivity to measurements performed on one of their subsystems 
and  only this subset of separable states is qualified to be classically correlated. 
However, OZ employed perfect orthogonal  projective measurements~\cite{OZ} 
on one of the subsystem  to quantify QC via  quantum discord, a measure proposed by them. 
 Henderson and Vedral~\cite{Ved} considered general measurements through local 
 positive operator valued measures (POVM) to separate pure CC from total correlations 
 in a bipartite state. Quantum deficit, another alternative measure proposed by Rajagopal and Rendell~\cite{RR}, 
determines genuine QC by finding how close a given quantum state is to its 
decohered classical counterpart. 

In this Letter, we address the basic issue of 
discerning quantumness of correlations, by introducing a novel theoretical measure  
{\em quantumness}, ${\cal Q}_{AB}$, where we employ  
  {\em generalized orthogonal projective measurements} on one part of a 
  bipartite state,  in an extended Hilbert space.  
In this new framework,  quantum discord~\cite{OZ} and quantum deficit~\cite{RR}, 
  are found to be  particular cases of quantumness. 
Separable states are shown to have zero {\em quantumness} - a significant feature not identified 
by the measures proposed earlier -   thus establishing 
   that absence of entanglement and classicality are synonymous.  
 Furthermore,  quantumness is shown to be (i) non-zero for all entangled states and 
 (ii) it provides an upper bound to  the relative entropy of entanglement of the bipartite state. 
This  provides the much needed connection between the results of measurement theory with 
those characterizing quantum entanglement. 
          
We begin with the considerations of OZ~\cite{OZ}, regarding  perfect 
measurements on the subsystem $A$ of a bipartite state 
$\hat\rho_{AB}$, defined by a set of one dimensional orthogonal projectors 
$\{\hat\Pi^A_\alpha\};\    \sum_\alpha\, \hat\Pi_\alpha^A=I_A$  {(\rm completeness)},
 $ \hat\Pi^A_\alpha\, \hat\Pi^A_{\alpha'}=\hat\Pi^A_\alpha\,\delta_{\alpha\, \alpha'}$ {(\rm orthogonality)},
where $\alpha$ distinguishes different outcomes of measurement.  The conditional density matrix
 of the subsystem $B$ - when measurement on $A$ is known to have led to the value $\alpha$  - is given by  
\begin{equation}
\label{rhoalpha}
\hat\rho^{(\alpha)}_{AB}=\frac{\hat\Pi^A_\alpha\otimes I_B\, \hat\rho_{AB}\, \hat\Pi^A_\alpha\otimes I_B}
{{\rm Tr}[\hat\Pi^A_\alpha\otimes I_B\, \hat\rho_{AB}]},
\end{equation} 
which occurs with a probability $p_\alpha={\rm Tr}[\hat\Pi^A_\alpha\otimes I_B\, \hat\rho_{AB}]$. The entropy 
$S(\hat\rho^{(\alpha)}_{AB})=-{\rm Tr}\, [\hat\rho^{(\alpha)}_{AB}\, \log\, 
\hat\rho^{(\alpha)}_{AB} ]$ of the state $\hat\rho^{(\alpha)}_{AB}$ of Eq.~(\ref{rhoalpha}) 
gives the uncertainty in interpreting the state of 
the system $B$, when the outcome $\alpha$ has been realized for the subsystem $A$. Given the results of the 
complete measurement 
$\{\hat\Pi^A_\alpha\},$ the conditional information entropy~\cite{positivity} is given by
\begin{equation}
\label{ce1}
S(B\vert A_{\{\hat\Pi^A_\alpha\}})=\displaystyle\sum_{\alpha}\, p_\alpha\, S(\hat\rho^{(\alpha)}_{AB}). 
\end{equation}  
On the other hand, a structural generalization of Shannon conditional information 
$H(B\vert A)=\sum_{a,b}P(a,b)\log P(a\vert b)=H(A,B)-H(B)$
(which is a consequence of the Bayes rule $P(b\vert a)=P(a,b)/P(a)$ for classical conditional probabilities) 
in terms of von Neumann entropies leads to an equivalent expression for the conditional entropies i.e.,  
{\scriptsize\begin{equation}
\label{ce2}
S(B\vert A)=S(A,B)-S(A)
=-{\rm Tr}\,[\hat\rho_{AB}\, \log\, \hat\rho_{AB}]-{\rm Tr}\,[\hat\rho_{A}\, \log\, \hat\rho_{A}]. 
\end{equation}}
{\em Quantum discord}~\cite{OZ} is defined 
 as the minimum value of the difference of two classically identical expressions Eqs. (\ref{ce1}) and 
(\ref{ce2}) for conditional entropies:
 \begin{equation}
 \label{discord}
 \delta(A,B)_{\{\hat\Pi^A_\alpha\}}={\rm min}_{\{\hat\Pi^A_\alpha\}}\, S(B\vert A_{\{\hat\Pi^A_\alpha\}})
 -S(B\vert A).
 \end{equation} 
(Here, the minimization  is taken over the set ${\{\hat\Pi^A_\alpha\}}$ of all orthogonal projectors, and it 
intends 
to find a scheme, which leaves the overall state of the system with 
{\em least} disturbance after measurement). Quantum discord vanishes iff  
 $S(B\vert A_{\{\hat\Pi^A_\alpha\}})=S(B\vert A);$ in such a situation, the correlations are  
well within the domain of Bayes rule of probability theory and are therefore {\em classical}. Also, 
{\scriptsize\begin{equation}
\delta(A,B)_{\{\hat\Pi^A_\alpha\}}=0\Leftrightarrow   
\hat\rho^{\rm (classical)}_{AB}= 
\sum_\alpha \hat\Pi^A_\alpha\otimes I_B\, \hat\rho^{\rm (classical)}_{AB}\, \hat\Pi^A_\alpha\otimes I_B,
\end{equation}} 
i.e., an optimal choice of measurement  $\{\hat\Pi^A_\alpha\}$ on $A$ leaves  
the overall state $\hat\rho^{\rm classical}_{AB}$ of a classically correlated system  unperturbed.  Quantum 
discord does not 
necessarily vanish for all separable states of the form Eq.~(\ref{wersep}) and this strongly suggests that   
{\em genuine QC imply more than entanglement.}    
A natural question would be to verify {\em if this identification is true, even when we employ more general 
measurements -
than the ones considered by OZ}.  

\noindent {\em Quantumness of correlations ${\cal Q}_{AB}$:} Let us consider 
the set of all tripartite   density operators $\{\hat\rho_{A'AB}\}$ in an extended Hilbert space 
${\cal H}_{A'}\otimes{\cal H}_A\otimes{\cal H}_B$, such that the bipartite 
state $\hat\rho_{AB}$ under investigation is a marginal of this extended system:    
${\rm Tr}_{A'}[\hat\rho_{A'AB}]=\hat\rho_{AB}.$   Carrying out 
an orthogonal projective measurement $\Pi_{i}^{(A'A)}; i=1,2,\ldots , $ 
on one of the subsystems $A'A$ of the three party state $\hat\rho_{A'AB}$, we find that     
{\scriptsize\begin{eqnarray}
\hat\rho_{A'AB}&\rightarrow &\hat\rho^{(i)}_{A'AB}=\frac{1}{p_i}\,\left[\hat\Pi^{(A'A)}_i\otimes I_B\, 
\hat\rho_{A'AB}\, 
\hat\Pi^{(A'A)}_i\otimes I_B\right]
  \\ 
{\rm and \ \ }\hat\rho_{AB}&\rightarrow &\hat\rho^{(i)}_{AB}=\frac{1}{p_i}\,{\rm 
Tr}_{A'}\left[\hat\Pi^{(A'A)}_i\otimes I_B\, 
\hat\rho_{A'AB}\, \hat\Pi^{(A'A)}_i\otimes I_B\right]  \nonumber
\end{eqnarray}}
with {\scriptsize$p_i~=~{\rm Tr}_{A'AB}~[\hat\Pi^{(A'A)}_i~\otimes~I_B~\hat\rho_{A'AB}]$} denoting the  
probability of occurrence.   
We define {\em Quantumness} ${\cal Q}_{AB}$ associated with a bipartite state 
$\hat\rho_{AB}$  as the  minimum Kullback-Liebler relative entropy~\cite{NC}
{\scriptsize\begin{eqnarray}
\label{qab}
&{\cal Q}_{AB}={\rm min}_{\{\hat\Pi^{(A'A)}_i,\, \hat\rho_{A'AB}\}}\,   S(\hat\rho_{AB}\vert\vert 
\hat\rho'_{AB})\hskip 1in \nonumber \\
&  \ \ = {\rm min}_{\{\hat\Pi^{(A'A)}_i,\, \hat\rho_{A'AB}\}}\left({\rm Tr}\, [\hat\rho_{AB}\log \hat\rho_{AB}]-
{\rm Tr}\, [\hat\rho_{AB}\log \hat\rho'_{AB}]\right),
\end{eqnarray}} 
where $\hat\rho'_{AB}$ denotes the residual state of the bipartite system,  after  the generalized measurement 
is performed:
$\hat\rho'_{AB}={\rm Tr}_{A'}\,[\sum_i\, \hat\Pi^{(A'A)}_i\otimes I_B\, 
\hat\rho_{A'AB}\, \hat\Pi^{(A'A)}_i\otimes I_B].$
 The minimum in Eq.~(\ref{qab}) is taken over the set $\{\hat\Pi^{(A'A)}_i\}$ of  projectors
 on the subsystems $A'A$ of all possible extendend states  $\{\hat\rho_{A'AB}\}$, which contain 
 $\hat\rho_{AB}$ as their marginal system.  
 
  The quantumness, ${\cal Q}_{AB}\geq 0$ (by definition), for all generalized measurements  - the equality sign 
holding iff 
 $\hat\rho'_{AB}=\hat\rho_{AB}$ i.e.,  quantumness vanishes iff the bipartite state $\hat\rho_{AB}$ 
 remains insensitive to generalized measurement  $\{\hat\Pi^{(A'A)}_i\}$. 
 It is easy to find that with every separable state of the form Eq.~(\ref{wersep}), there exists a 
 tripartite density operator $\hat\rho_{A'AB}=\sum_{i}\, p_i \hat\Pi_i^{(A'A)}\otimes \hat\rho^{(i)}_B$, with  
 ${\rm Tr}_{A'}\, [\hat\Pi_i^{(A'A)}]=\hat\rho^{(i)}_A,$ so that ${\rm 
Tr}_{A'}\,[\hat\rho_{A'AB}]=\hat\rho^{{\rm(sep)}}_{AB}=\sum_i\, p_i\, 
\hat\rho^{(i)}_A~\otimes~\hat\rho^{(i)}_B.$ Clearly, a separable  state is left unperturbed under the 
generalized measurements
  $\{\hat\Pi^{(A'A)}_i\}$ on one end (i.e, on the subsystem $A'A$) of the tripartite state $\hat\rho_{A'AB}$ 
i.e.,  
  $\hat\rho^{{\rm (sep)}'}_{AB}=\hat\rho^{{\rm (sep)}}_{AB}$. Hence, 
 quantumness ${\cal Q}_{AB}$ vanishes for all  separable states.
This is a striking aspect - not identified by the measures proposed previously - 
which establishes that separability and classicality imply each other. 

We illustrate this with an example. Consider a separable mixture of non-orthogonal states of two qubits,  
$\hat\rho_{AB}=p\, \vert 0_A,0_B\rangle\, \langle 0_A,0_B\vert + (1-p)\,  \vert +_A,+_B\rangle\, \langle 
+_A,+_B\vert$
(with $\vert \pm\rangle = \frac{1}{\sqrt{2}}\,(\vert 0\rangle\pm\vert 1\rangle),$ and  $0\leq p\leq 1$), which 
is found to have 
non-classical correlations~\cite{Ved, Hamieh}. Following our procedure, a three qubit state 
$\hat\rho_{A'AB}=p\, \vert 1_{A'},0_A,0_B\rangle\, \langle 1_{A'}, 0_A,0_B\vert + (1~-~p)\,  
\vert 0_{A'}, +_A,+_B\rangle\, \langle 0_{A'},+_A,+_B\vert$ 
may be constructed such that ${\rm Tr}_{A'}[\hat\rho_{A'AB}]=\hat\rho_{AB}$. An optimal measurement on  $A'A$ 
constitues a set of complete, orthogonal 
projectors {\scriptsize$\{\hat\Pi_{1}^{(A'A)}=\vert 1_{A'},0_A\rangle\, \langle 1_{A'},0_A\vert, \ 
\hat\Pi_{2}^{(A'A)}=\vert 1_{A'},1_A\rangle\, \langle 1_{A'},1_A\vert ,$}
{\scriptsize$\hat\Pi_{3}^{(A'A)}=\vert 0_{A'},+_A\rangle\, \langle 0_{A'},+_A\vert,$}  
{\scriptsize$\hat\Pi_{4}^{(A'A)}=\vert 0_{A'},-_A\rangle\, \langle 0_{A'},-_A\vert  \}$}, and this leaves the 
overall state unperturbed: 
{\scriptsize$\sum_{i=1}^{4}\, \hat\Pi^{(A'A)}_i~\otimes~I_B\, \hat\rho_{A'AB}\, 
\hat\Pi^{(A'A)}_i~\otimes~I_B=\hat\rho_{A'AB},$} 
and {\scriptsize$\hat\rho'_{AB}=\hat\rho_{AB}.$} Hence, {\scriptsize${\cal Q}_{AB}=0$} 
for this state~\cite{HS}.    

There arises a question on the operational aspects of finding an optimal measurement 
over the {\em set of  all extended states} $\{\hat\rho_{A'AB}\}$, as we have access only to the marginal 
bipartite state 
$\hat\rho_{AB}.$ It may be noted that when the minimization in Eq.~(\ref{qab}) is done by confining ourselves 
  to three party extended systems of the direct product form $\hat\rho_{A'}\otimes\hat\rho_{AB},$ the  
orthogonal 
  projective measurement $\{\hat\Pi^{(A'A)}_{i}\}$ corresponds to a set of POVMs 
  $\{V_i^{A}\};\ \sum_i V_i^{A\dag} V_i^{A}=I_A,$ on the subsystem $A$ of the bipartite state 
$\hat\rho_{AB},$ transforming it to,  $\hat\rho'_{AB}=\sum_{i}\, V_i^{A}\otimes I_B\, \hat\rho_{AB}\, 
V_i^{A\dag}\otimes I_B$, 
as a result of generalized measurement
and no explicit reference on the extended state $\hat\rho_{A'}\otimes\hat\rho_{AB}$ is needed 
in such situations. There still remains a problem concerning optimization involving  
more general extended states $\hat\rho_{A'AB},$ that do not have 
the direct product structure $\hat\rho_{A'}\otimes\hat\rho_{AB};$ but this will be resolved at the end of this 
Letter, where we prove that   
quantumness ${\cal Q}_{AB}={\rm min}_{\{\hat\rho_{AB}^{{\rm (sep)}}\}}\, S(\hat\rho_{AB}\vert\vert  
\hat\varrho^{{\rm (sep)}}_{AB}),$ 
where $\{\hat\varrho_{AB}^{{\rm (sep)}}\}$ is the  set of all bipartite separable states sharing the same 
marginal system $\hat\rho_{B}$  for the part which does not come under the action of 
generalized measurements. 

\noindent{\em Unifying features of quantumness:} 
 We will now show that previously proposed measures of QC are special cases of $Q_{AB}$.  
 First, let us  consider  quantum deficit~\cite{RR} $D_{AB}$ of a bipartite state $\hat\rho_{AB}$ given by,
{\scriptsize\begin{equation}
 \label{dab}
 D_{AB}=S(\hat\rho_{AB}\vert\vert\hat\rho^{(d)}_{AB})
 = {\rm Tr}\, [\hat\rho_{AB}\log\hat\rho_{AB}]-{\rm Tr}\, [\hat\rho_{AB}\log\hat\rho^{d}_{AB}],  
 \end{equation}}
 where  $\rho^{(d)}_{AB}$ corresponds to the classical (decohered) density operator 
  with the same marginal states~\cite{Partovi} $\hat\rho_A,\ \hat\rho_B$ 
  as that of $\hat\rho_{AB}$.  The quantum deficit $D_{AB}$ determines the quantum excess of correlations in the 
state 
    $\hat\rho_{AB}$, with reference to its classical counterpart  $\hat\rho^{(d)}_{AB}.$  
The decohered state $\hat\rho^{(d)}_{AB}=\sum_{a,b}\, P(a,b)\,\vert a\rangle\, 
\langle a\vert   \otimes \vert b\rangle\, \langle b\vert$ is diagonal in the eigenbasis 
	$\{\vert a\rangle\}, \ \{\vert b\rangle\}$ of the marginal density operators $\hat\rho_A,$  $\hat\rho_B$ 
	(with $P(a,b)=\rho_{a,b;a,b}$ denoting the diagonal elements of $\hat\rho_{AB}$, in the eigen basis of its 
subsystems~\cite{RR}) and 
	    is easy to see that if the optimal measurements that minimize the quantumness~(see Eq.~(\ref{qab})) 
	   of the bipartite state $\hat\rho_{AB}$ are the eigen projectors 
	   $\{\hat\Pi_a^{A}=\vert a\rangle\, \langle a\vert\}$ of the subsystem $A$ - with no need to invoke  
measurements 
	   in an extended space - quantumness ${\cal Q}_{AB}$ is equal to the quantum deficit $D_{AB}.$ 
	   
	   Consider bipartite states $\hat\rho_{AB}$, which allow  
  optimization of  ${\cal Q}_{AB}$ in terms of  orthogonal projective set $\{\hat\Pi_{\alpha}^{A}\}$ on the 
subsystem $A$
   (i.e., there is no need to invoke extended states $\hat\rho_{A'AB}$, if minimization of quantumness 
   could be acheived using actual space perfect orthogonal measurements). Using (i) the completeness  
$\sum_\alpha\, \hat\Pi_\alpha^A~\otimes~I_B=I_A\otimes I_B$, (ii) the orthogonality  
$\hat\Pi_\alpha^A\otimes I_B\hat\Pi_{\alpha'}^A\otimes I_B=\hat\Pi_\alpha^A\otimes I_B\, 
\delta_{\alpha,\alpha'},$
and  (iii) the property $ [\hat\Pi_{\alpha}^{A}\otimes I_B,(\hat\rho'_{AB})^k]=0;\ k=1,2,\ldots$ (which follows 
readily from 
the structure of the bipartite state $\hat\rho'_{AB}=\sum_\alpha\,  \hat\Pi_{\alpha}^{A}\otimes I_B\, 
\hat\rho_{AB}\hat\Pi_{\alpha}^{A}\otimes I_B$ 
 after the projective measurement $\{\hat\Pi_{\alpha}^{A}\}$),  
it may be seen that  
{\scriptsize\begin{eqnarray}
{\rm Tr}\,[\hat\rho_{AB}\log \hat\rho'_{AB}]&=& {\rm Tr}\,[\sum_{\alpha}
\, \hat\Pi_\alpha^A\otimes I_B\,\hat\rho_{AB}\log \hat\rho'_{AB}],\nonumber \\ 
&=& {\rm Tr}\,[
\sum_\alpha\, \hat\Pi_\alpha^A\otimes I_B\,\hat\rho_{AB}\log \hat\rho'_{AB}\,\hat\Pi_\alpha^A\otimes I_B], 
\nonumber \\ 
&=& {\rm Tr}\,[\sum_\alpha\, \hat\Pi_\alpha^A\otimes I_B\,\hat\rho_{AB}\,\hat\Pi_\alpha^A\otimes I_B\,\log 
\hat\rho'_{AB}], 
\nonumber \\ 
&=& {\rm Tr}\,[\hat\rho'_{AB}\log \hat\rho'_{AB}].
\end{eqnarray}} 
In this particular case, the quantumness assumes the form,   
{\scriptsize\begin{equation}
\label{sqab}
{\cal Q}_{AB}={\rm min}_{\{\Pi_\alpha^A\}}\,S(\hat\rho_{AB}\vert\vert\hat\rho_{AB}')=
{\rm min}_{\{\Pi_\alpha^A\}}[S(\hat\rho'_{AB})-S(\hat\rho_{AB})],
\end{equation}}
which immediately leads to~\cite{proposition}
\begin{equation}
{\cal Q}_{AB}= \delta(A,B)_{\{\Pi^A_\alpha\}}+ S(\hat\rho_{A}\vert\vert 
\hat\rho'_{A})
\end{equation}
 relating quantumness and quantum discord. 
Further, when the optimal measurements  correspond to the set of  eigenprojectors  $\{\Pi^A_a\}$ of the 
marginal density matrix $\hat\rho_A$ (in which case $\hat\rho_A$ coincides with $\hat\rho'_A$ and hence, 
$S(\hat\rho_{A}\vert\vert 
\hat\rho'_{A})=0$ ), all the three measures 
${\cal Q}_{AB},$ $\delta(A,B)_{\{\Pi^A_\alpha\}}$ and $D_{AB}$ of quantum correlations are identical: 
\begin{equation}
\label{equal}
{\cal Q}_{AB}=\delta(A,B)_{\{\Pi^A_a\}}=D(A,B).
\end{equation}
An explicit example of two qubits, $\hat\rho_{AB}=p\, \vert\phi_{AB}^{+}\rangle\langle \phi_{AB}^{+}\vert  
+(1-p)\, 
\vert\phi^{-}_{AB}\rangle\langle\phi^{-}_{AB}\vert$ (with  $\vert\phi_{AB}^{\pm}\rangle=
\frac{1}{\sqrt{2}}[\vert 0_A,0_B\rangle\pm \vert 1_A,1_B\rangle],$ denoting the Bell states and 
$0\leq p\leq 1$) demonstrates the above unifying feature.  The quantum deficit for this state is given by  
$D_{AB}=\log 2+p\log p+(1-p)\,\log (1-p)$. On the other hand, quantum discord is optimized~\cite{OZ} by 
the set of orthogonal projectors $\{\hat\Pi^{A}_1=\vert 0_A\rangle\langle 0_A\vert, \hat\Pi^{A}_2
=\vert 1_A\rangle\langle 1_A\vert \}$ and we obtain $\delta(A,B)_{\{\Pi^A_a\}}=D_{AB}$ in this case. 
It may be noted that the quantum deficit and discord are equal to the relative entropy of 
entanglement $E_{{\rm RE}}(\hat\rho_{AB})$ in this particular example~\cite{Ved, PlVed2}. 
The quantumness of correlations is also equal to  the relative entropy of entanglement in this case, 
which  follows from a more general result ${\cal Q}_{AB}\geq E_{{\rm RE}}(\hat\rho_{AB})$ between the two as 
established below.

\noindent{\em The quantumness and the relative entropy of entanglement:} 
Let us express the extended three party density operator $\hat\rho_{A'AB}$ 
in terms of a complete, orthogonal set of basis states 
$\{\vert j_{A'A}\rangle \otimes \vert \beta_B\rangle\}$ of subsystems $A'A$ and $B$ as, 
$\hat\rho_{A'AB}=\displaystyle\sum_{j',j, \beta',\beta}\, \rho_{j',\beta'; j,\beta}\, 
\vert j'_{A'A}\rangle\, \langle j_{A'A}\vert  \otimes \vert \beta'_B\rangle\, \langle \beta_B\vert.$
Corresponding to a measurement $\hat\Pi_{i}^{(A'A)}=\vert i_{A'A}\rangle\, \langle i_{A'A}\vert$ on $A'A,$ the 
state gets 
projected to,
{\scriptsize\begin{equation}
\label{rho'}
\frac{\hat\Pi^{(A'A)}_i\otimes I_B\, \hat\rho_{A'AB}\, 
\hat\Pi^{(A'A)}_i\otimes I_B}{{\rm Tr}\,  [\hat\Pi^{(A'A)}_i~\otimes~I_B~\,~\hat\rho_{A'AB}]}=
\displaystyle\sum_{\beta',\beta}\, \frac{\rho_{i,\beta'; i,\beta}}{p_i}\, \hat\Pi_i^{(A'A)}\otimes 
\vert\beta'_B\rangle\,\langle\beta_B\vert 
\end{equation}}
(with {\scriptsize$p_i={\rm Tr}\,  
[\hat\Pi^{(A'A)}_i~\otimes~I_B~\,~\hat\rho_{A'AB}]=\sum_{\beta}\,\rho_{i,\beta;i,\beta}$}) 
and after carrying out the complete measurement $\{\hat\Pi_{i}^{(A'A)}\}$, 
the marginal bipartite system $\hat\rho_{AB}$ is transformed into, 
{\scriptsize\begin{eqnarray}
\label{sep2}
\hat\rho'_{AB}&=& {\rm Tr}_{A'}\, [\sum_i\, 
\hat\Pi^{(A'A)}_i\otimes I_B\, \hat\rho_{A'AB}\, 
\hat\Pi^{(A'A)}_i\otimes I_B] \nonumber \\
 &=& \displaystyle\sum_i\, p_i\, \hat\rho^{(i)}_A\otimes \hat\varrho^{(i)}_{B},
\end{eqnarray}}
where (see Eq.~(\ref{rho'})) 
 {\scriptsize$\hat\rho^{(i)}_A={\rm Tr}_{A'}\,[\hat\Pi^{(A'A)}_i],$}     
{\scriptsize$\hat\varrho^{(i)}_B=\sum_{\beta',\beta}\,\frac{\rho_{i,\beta';i,\beta}}{p_i}\, 
\vert\beta'_B\rangle\,\langle\beta_B\vert,$} with
 {\scriptsize $\sum_i p_i\, \hat\varrho^i_B =\hat\rho_B={\rm Tr}_{A}\,[\hat\rho'_{AB}].$}
From Eq.~(\ref{sep2}), it is obvious that the state $\hat\rho'_{AB}$ of the bipartite system, left 
after performing the generalized measurement $\{\hat\Pi^{(A'A)}_i\}$ on a part  
of the system, is a separable state. As the optimization of quantumness ${\cal Q}_{AB}$ is done over the set of 
all projectors $\{\hat\Pi^{(A'A)}_i\},$ and the set of  all   extended states $\{\hat\rho_{A'AB}\}$, it is easy 
to see 
that   $\{\hat\rho'_{AB}=\hat\varrho^{{\rm (sep)}}_{AB};\hat\rho_B={\rm Tr}[\hat\rho'_{AB}] \}$
corresponds to {\em the set of  all separable states sharing the same subsystem density matrix $\hat\rho_B$} for 
the part $B$, which does not come under the direct action 
of generalized measurements $\{\hat\Pi^{(A'A)}_i\}$. Thus, we obtain 
\begin{eqnarray}
\label{relent}
{\cal Q}_{AB}&=&{\rm min}_{\{\hat\Pi^{(A'A)}_i,\, \hat\rho_{A'AB}\}}\,   S(\hat\rho_{AB}\vert\vert 
\hat\rho'_{AB})\nonumber \\ 
&=& {\rm min}_{\{\hat\varrho_{AB}^{{\rm (sep)}}\}}\, S(\hat\rho_{AB}\vert\vert \hat\varrho^{{\rm (sep)}}_{AB})
\end{eqnarray} 
with minimization taken over the set of all separable states $\{\hat\varrho^{{\rm (sep)}}_{AB}; \hat\rho_B={\rm 
Tr}[\hat\varrho^{\rm (sep)}_{AB}]\}.$ In other words,  quantumness ${\cal Q}_{AB}$ is the 
minimum distance of the bipartite state $\hat\rho_{AB}$ with the set of all separable states 
$\{\hat\varrho^{{\rm (sep)}}_{AB};\ {\rm Tr}\,[\hat\varrho^{{\rm (sep)}}_{AB}]=\hat\rho_B\}.$ 
From (\ref{relent}) it is evident that quantumness ${\cal Q}_{AB}$ is 
necessarily non-zero for all entangled bipartite states $\hat\rho_{AB}$ and it serves as an upper bound to the   
the relative entropy of entanglement $E_{\rm RE}(\hat\rho_{AB})$ ~\cite{PlVed2}. 
This is a significant theoretical result as it   
provides a natural link between  the concepts of quantum entanglement with 
those based on quantum measurement theory. Most importantly, this  establishes a flawless merging of {\em 
quantumness of correlations} with {\em quantum entanglement} itself.    

\noindent{\em Separating total correlations into classical and quantum parts:} 
In classical information theory, Shannon mutual information is an unequivocal measure of correlations. 
Its quantum analogue, the von Neumann mutual information,  $S(A:B)=S(\hat\rho_{AB}\vert\vert \hat\rho_A\otimes 
\hat\rho_B)$ 
quantifies the {\em total} correlations in a bipartite state $\hat\rho_{AB},$ with 
$\hat\rho_A,\ \hat\rho_B$ denoting the subsystem density operators. HV~\cite{Ved} proposed that 
  {\scriptsize$C_A(\hat\rho_{AB})={\rm max}_{\{V^A_i\}}\, S(\hat\rho_B)-\sum_i\, q_i S(\hat\rho^i_B),$}
-  the residual information entropy of $B$, after carrying out a POVM measurement {\scriptsize$\{V^A_i\}$} 
 on the subsystem $A$ (with  \ {\scriptsize$\hat\rho^i_B={\rm Tr}_{A}[V^A_i\otimes I_B\, \hat\rho_{AB}\, 
V_i^{A\dag}\otimes
I_B]/q_i;$}  {\scriptsize$q_i={\rm Tr}_{AB}[V^A_i\otimes I_B\, \hat\rho_{AB}\, V_i^{A\dag}\otimes I_B]$})  - 
serves as a measure of CC. 
By analyzing some examples they found that classical and entangled correlations do not add up to give  
 total correlations~\cite{Ved}: {\scriptsize $C_A(\hat\rho_{AB})+E_{\rm RE}(\hat\rho_{AB})\neq S(A:B).$} 
 This leaves open the question, 
 {\em ``Are different types of correlations  not additive?''} We find that  
 generalized projective measurements {\scriptsize$\{\hat\Pi^{(A'A)}_{i}\}$} on extended three party states of 
the product form, 
{\scriptsize$\hat\rho_{A'AB}=\hat\rho_{A'}~\otimes~\hat\rho_{AB}$} 
lead to,  {\scriptsize${\rm min}_{\{\hat\Pi^{(A'A)}_{i}, \hat\rho_{A'}\otimes\hat\rho_{AB}\}}
\,S(\hat\rho_{A'}\otimes\hat\rho_{AB}\vert\vert 
\hat\rho'_{A'AB})+\sum_i q_i\log q_i$}
 {\scriptsize $\leq S(A:B)-C_A(\hat\rho_{AB})$}, giving us 
 {\scriptsize$ {\rm min}_{\{\hat\Pi^{(A'A)}_i, 
\hat\rho_{A'}\otimes\hat\rho_{AB}\}}\,S(\hat\rho_{AB}\vert\vert \hat\rho'_{AB})\leq 
S(A:B)-C_A(\hat\rho_{AB})-\sum_i q_i\log q_i.$} And, as  
 {\scriptsize$E_{\rm RE}(\hat\rho_{AB})\leq{\cal Q}_{AB}\leq {\rm min}_{\{\hat\Pi^{(A'A)}_i,\hat\rho_{A'}\otimes 
\hat\rho_{AB}\}}\,S(\hat\rho_{AB}\vert\vert \hat\rho'_{AB}),$} 
 we obtain  the  inequality $C_A(\hat\rho_{AB})+E_{\rm RE}(\hat\rho_{AB})+\sum_i q_i\log q_i \leq S(A:B).$
On the other hand, a consistent separation of total correlations  into classical and quantum correlations as 
$S(A:B)={\cal C}_{AB}+{\cal Q}_{AB}$   
would require a generalization of the HV measure $C_A(\hat\rho_{AB})$ of CC as, 
\begin{equation} 
 {\cal C}_{AB}=S(\hat\rho_{AB}\vert\vert\hat\rho_A\otimes\hat\rho_B)-{\rm min}_{\{\hat\Pi^{(A'A)}_i,\, 
\hat\rho_{A'AB}\}}\,S(\hat\rho_{AB}\vert\vert \hat\rho'_{AB})
\end{equation}           
with the set $\{\hat\rho_{A'AB}\}$ exhausting  all possible three party extensions of the quantum state 
$\hat\rho_{AB}$ - there being no restriction on the product structure, $\hat\rho_{A'AB}=\hat\rho_{A'}\otimes 
\hat\rho_{AB}.$   

In conclusion, this is a contribution towards resolving the difficulties  
faced in quantifying the residual quantum correlations in bipartite separable states,
through the  introduction of a novel measure called quantumness, ${\cal Q}_{AB}$. 
This brings out the need for the crucial inclusion 
of a generalized measurement scheme in quantifying quantum correlations. 
Separable states are insensitive to an optimal generalized measurement and  
are shown to have zero quantumness  - an important feature, which the measures 
proposed previously failed to recognize.  An 
entangled bipartite state gets projected into its 
closest separable product form, which shares the same marginal for one of its subsystems - consequent to an 
optimal  measurement on the other part -  showing that 
quantumness of correlations is non-zero for all  entangled states and it serves as an upper bound to the 
relative entropy of entanglement of the bipartite state. 

We thank Dr. Ron Rendell for drawing our attention to Ref.~\cite{Hamieh} and Dr. Dong Yong for pointing out some 
subtle technical details, which we had missed.


\begin{thebibliography}{1}
\bibitem{positivity} Note that the conditional entropy $S(B\vert A_{\{\hat\Pi^{A}_\alpha\}})$ is (i)  positive  
and 
(ii) it exhibits an inherent measurement dependence.
\bibitem{HS}  The extended three party state $\hat\rho_{A'AB}\neq \hat\rho_A'\otimes\hat\rho_{AB},$ 
and so, the orthogonal projective measurement $\{\Pi_{i}^{A'A}\}$ can not be expressed as a POVM transformation 
  on the state $\hat\rho_{AB}.$  A POVM optimization  had shown~\cite{Ved,Hamieh} that the state exhibits 
quantum correlations.
\end{thebibliography}
\end{document}